# Damped Lyman Alpha Systems vs. Cold + Hot Dark Matter

**Anatoly Klypin** (aklypin@nmsu.edu)

Department of Astronomy, New Mexico State University

Las Cruces, NM 88001

**Stefano Borgani** (borgani@vaxpg.pg.infn.it)

INFN, Sezione di Perugia, c/o Dp. di Fisica dell'Università

via A. Pascoli, I-06100 Perugia, Italy

and

International School for Advanced Studies

Via Beirut 2-4, I–34014 Trieste, Italy

**Jon Holtzman** (holtz@lowell.edu)

Lowell Observatory, Mars Hill Road, Flagstaff, AZ 86100

**Joel Primack** (joel@lick.ucsc.edu)

Physics Department, University of California, Santa Cruz, CA 95064

**Abstract.** Although the Cold + Hot Dark Matter (CHDM) cosmology provides perhaps the best fit of any model to all the available data at the current epoch, CHDM produces structure at relatively low redshifts and thus could be ruled out if there were evidence for formation of massive objects at high redshifts. Damped Ly$\alpha$ systems are abundant in quasar absorption spectra and thus provide possibly the most significant evidence for early structure formation, and thus perhaps the most stringent constraint on CHDM. Using the numbers of halos in N-body simulations to normalize Press-Schechter estimates of the number densities of protogalaxies as a function of redshift, we find that CHDM with $\Omega_c/\Omega_\nu/\Omega_b = 0.6/0.3/0.1$ is compatible with the damped Ly$\alpha$ data at $\leq 2.5$, but that it is probably incompatible with the limited $z > 3$ damped Ly$\alpha$ data. The situation is uncertain because there is very little data for $z > 3$, and also it is unclear whether all damped Ly$\alpha$ systems are associated with collapsed protogalaxies. The predictions of CHDM are quite sensitive to the hot (neutrino) fraction, and we find that $\Omega_c/\Omega_\nu/\Omega_b = 0.675/0.25/0.075$ is compatible even with the $z > 3$ data. This corresponds to lowering the neutrino mass from 6.8 to 5.7 eV, for $H_0 = 50$ km s$^{-1}$ Mpc$^{-1}$. In CHDM, the higher redshift damped Ly$\alpha$ systems are predicted to have lower masses, which can be checked by measuring the velocity widths of the associated metal line systems.

*Subject headings:* cosmology: theory — dark matter — large-scale structure of universe — galaxies: formation — quasars: absorption lines

# 1. Introduction

The standard Cold Dark Matter (CDM) cosmology is based on the assumptions that structure grew by gravitational collapse from initial adiabatic Gaussian fluctuations with a Harrison-Zel'dovich spectrum in a critical-density (i.e. $\Omega = 1$) universe as predicted by simple cosmic inflation models, together with the assumption that the dark matter is "cold" – i.e., fluctuations of all cosmologically interesting sizes are preserved during the radiation era (Blumenthal et al. 1984). Although CDM with bias parameter $b \approx 2$ was found to be consistent with data from galaxy to cluster scales, it appears to be less compatible with data on larger scales (see e.g. Davis et al. 1992). In particular, if cosmological theories are normalized to produce cosmic microwave background temperature anisotropies compatible with the COBE measurements (Smoot et al. 1992, Bennett et al. 1994), then Cold Dark Matter (CDM) produces somewhat too much structure on scales of several 10's of Mpc (e.g., too many clusters; cf. White, Efstathiou, & Frenk 1993) and far too much structure on scales of a few Mpc (e.g., median rms velocities in groups are far higher than observations; cf. Nolthenius et al. 1994).

Cold + Hot Dark Matter (CHDM) is based on exactly the same assumptions as standard CDM, except that in addition to the cold dark matter it is assumed that one of the three known species of neutrinos has a mass of $22.7\Omega_\nu h_{50}^2$ eV, where $\Omega_\nu$ is the fraction of critical density contributed by these neutrinos and $h_{50}$ is the Hubble parameter in units of 50 km s$^{-1}$ Mpc$^{-1}$. These neutrinos would have been highly relativistic at the era (approximately a year after the Big Bang) when the horizon first encompassed $10^{12}$ $M_\odot$, since the temperature then was ~keV; thus they would be "hot" dark matter. Their free streaming would have erased fluctuations in their density on scales from galaxies through clusters during the radiation era, and their still-significant random velocities at more recent epochs would have impeded their clustering on small scales. Consequently, if the CDM and CHDM density fluctuation spectra $P(k)$ are both normalized to agree with COBE, as in Figure 1(a), then CHDM has a little less power on intermediate scales and much less power on small scales (i.e., large $k = 2\pi/\lambda$). Approximate linear calculations (Holtzman 1989, Holtzman & Primack 1993; cf. also Schaefer, Shafi, & Stecker 1989; van Dalen & Schaefer 1992; Schaefer & Shafi 1992, 1994; Taylor & Rowan-Robinson 1992; Wright et al. 1992) showed that the most promising parameters for CHDM were $\Omega_c/\Omega_\nu/\Omega_b \approx 0.6/0.3/0.1$. Cosmological N-body simulations have confirmed that CHDM with this combination of parameters appears to be compatible with all the available data at the current epoch (Klypin et al. 1993 [KHPR93] and references therein; in disagreement with claims of Cen & Ostriker 1994, Nolthenius et al. 1994 [NKP94] show that velocities predicted by CHDM are compatible with observations on $\sim 1$ $h^{-1}$ Mpc scales [where as usual $h$ is the Hubble parameter in units of 100 km s$^{-1}$ Mpc$^{-1}$], and Klypin & Rhee 1994 and Jing & Fang 1994 show that CHDM predictions are compatible with observed cluster properties).

It should be understood that the linear calculations were done for a fairly widely spaced grid of values of parameters such as $\Omega_\nu$, and that better fits could possibly be found for intermediate values. Holtzman, Klypin, & Primack (1994) have done a new set of linear calculations, including many intermediate values of parameters (some examples are given in Figure 1(a)), in order to help to choose parameters for new high-resolution N-body simulations. Note that for long waves ($k < 1h$ Mpc $^{-1}$) the difference in power between the standard 0.6/0.3/0.1 CHDM model and these variants is small, which insures that all large scale properties (e.g. bulk velocities, number of clusters, and the cluster-cluster correlation function) will be almost unchanged. As we go to smaller scales the difference becomes larger and is as much as 50% on galactic scales. This is already enough to increase significantly the number of galaxies at high redshifts. Another parameter that can affect the spectrum of fluctuations is $\Omega_b$, the contribution of baryons to the mean density. Two sorts of observations tend to pull it in opposite directions. Abundances of light nuclides, especially $^4$He, together with standard Big Bang Nucleosynthesis (BBN), give low estimates: $\Omega_b \sim 0.05$, with $\Omega_b \sim 0.1$ being marginally consistent with the latest observational results (T. P. Walker, private communication). At the same time, results from x-ray studies of galaxy clusters favor higher values of $\Omega_b$ (White et al. 1993). Since we think that estimates of the masses of different components of clusters are still not as reliable as the constraints from BBN, we choose $\Omega_b$ higher than the usual BBN value of $0.05h_{50}^2$, but still compatible with BBN. Note that lowering $\Omega_b$ at constant $\Omega_\nu$ increases the amplitude of fluctuations at small scales while the amplitude at large scale is unaffected.

In this paper, we consider the implications of CHDM for damped Ly$\alpha$ systems. Damped Ly$\alpha$ systems are observed as wide absorption troughs in quasar spectra, indicating that the line of sight to the quasar intersects a cloud of neutral hydrogen with a column density $\geq 10^{20}$ cm$^{-2}$ (for a recent review, see Wolfe 1993). Damped Ly$\alpha$ systems are sufficiently common that their abundance can be characterized statistically out to $z \approx 3$ (Lanzetta 1993, Lanzetta et al. 1993 [LWT93]). They thus provide potentially the most reliable and important information presently available on early structure formation. Three recent preprints (Subramanian & Padmanabhan 1994, Mo & Miralda-Escude 1994, and Kauffmann & Charlot 1994) have claimed that the statistics on damped Ly$\alpha$ systems are strongly incompatible with the predictions of CHDM.

There are some arguments suggesting that the damped Ly$\alpha$ systems correspond to massive protogalaxies. We note that the arguments typically have the sense of double negative statements: observational data are *not incompatible* with large spiral disks. For example, Briggs et al. (1989) found a high column density absorber at $z = 2.04$ with a complicated substructure: at least two or three components in projection; two components in the absorption spectrum with difference in velocities $\Delta v \sim 16$ km s$^{-1}$ and width $\sigma \sim 10$ km s$^{-1}$. These small velocities could be interpreted in favor of a small mass of the system. But Briggs et al. argue that this is likely a large galaxy with rotational velocity $v_{\rm los} \sim 200$ km s$^{-1}$ because a *spherical* model is incompatible with the data. We note that this is an interpretation of the results, which probably excludes only a very naive model. No direct evidence of large velocities was actually found. Thus, observationally, the situation is quite unclear. There are arguments that damped Ly$\alpha$ systems are gas-rich dwarf galaxies (Tyson 1988; Hunstead, Pettini, & Fletcher 1990). It is even possible that some of the damped Ly$\alpha$ systems observed are objects that have collapsed only in one or two dimensions (pancakes or caustics) rather than fully collapsed (proto)galaxies. Taking into account the uncertainties in both the interpretation of the observational results and in theoretical predictions about how galaxy progenitors should look at high redshifts, we will consider the possibility that damped Ly$\alpha$ clouds at higher redshifts are protogalaxies less massive than present day $L_*$ galaxies. In §2 we discuss the number density of dark matter halos of various masses as a function of redshift in CHDM, using N-body simulations to normalize Press-Schechter estimates. In §3, we compare the numbers of damped Ly$\alpha$ systems predicted by CHDM to the data of LWT93. §4 presents our conclusions – which are in disagreement with the recent preprints just mentioned. To paraphrase Mark Twain's cable to the Associated Press on reading his obituary in 1897, "The reports of my [model's] death are greatly exaggerated."

## 2. Abundance of Dark Halos

KHPR93 reported detailed results from CDM and CHDM N-body simulations with boxes of size 7, 25, and 100 $h^{-1}$ Mpc and PM force resolution of $256^3$; NKP94 reported preliminary results from simulations with boxes of size 50 $h^{-1}$ Mpc and force resolution of $512^3$, which are discussed in detail in Klypin, Nolthenius, & Primack (1994). In Figure 1(b) we compare the numbers of halos of masses $1.5 \times 10^{11}$, $5 \times 10^{11}$, and $1.5 \times 10^{12}$ $h^{-1} M_\odot$ from the KHPR93 and NKP94 simulations (shown as dashed and solid lines, respectively) with Press-Schechter (Press & Schechter 1974, Efstathiou & Rees 1988, Bond et al. 1992) estimates (dot-dashed lines). Halos in numerical simulations were identified as local maxima of the total density (hot + cold) on a mesh with $97.5h^{-1}$kpc cell size. The mass assigned to a halo was defined as the sum of masses inside the central cell and its nearest 26 neighbors (thus, a $3^3$ cube in space). The effective comoving radius is thus $0.181h^{-1}$Mpc, which gives proper radius $60h^{-1}$kpc at $z = 2$. The halos should have central overdensity of more than 50 to be considered as candidates for "galaxies". Because a halo is typically split between four or more cells, the limit roughly corresponds to a real overdensity $\approx 200$.

In order to check that and to gain further insight into the structure of the halos, we applied a more accurate, but more time consuming, algorithm to our 50 $h^{-1}$ Mpc CHDM$_2$ simulation (NKP94) at $z = 2.2$, and compared its results with our faster maxima-on-mesh approach. The new algorithm finds positions of spheres of given constant radius and gives the mass inside the spheres. The algorithm is an iterative procedure. At the beginning, it places a sphere of given radius at each maximum of density defined on the mesh. Then it finds the center of mass of all cold particles inside the sphere. Then the center of the sphere is displaced to the center of mass. This procedure is then repeated. The algorithm stops once either the displacement is smaller than 0.005 of the cell size or the mass inside the sphere stops rising. The comoving radius of the sphere was chosen to be $60h^{-1}$kpc or 0.61 of the cell size. Note that except for the initial location, this method does not involve the mesh (so there are no problems with the positions of halos relative to the mesh). Typically 4–6 iterations (with a maximum of 10-12) are needed to find the position that corresponds to a local maximum of cold mass inside the sphere.

Finally, we place three spheres of different radii centered on each halo found. The radii were different by a factor 2 from one another: $180h^{-1}$kpc (the effective radius of our maxima-on-mesh approach), $90h^{-1}$kpc, and $45h^{-1}$kpc. The diameter of the smallest sphere is about a cell size, which is marginally resolved. We found the following. (1) The mean density contrast of cold dark matter within the sphere increases by a factor 2–5 when its radius decreases by a factor of two. This is roughly consistent with the factor of four expected for an isothermal distribution with $\rho \propto r^{-2}$. (2) The mass within $180h^{-1}$kpc tightly correlates with the mass found with the maxima-on-mesh approach. The former is 20–25% higher and the spread is 10–20%. The difference is small and could be interpreted as if the effective radius for maxima-on-mesh should be $167h^{-1}$kpc, not $180h^{-1}$kpc. The difference is not very significant. Because of the steep growth of the density toward the centers of the dark halos in the simulations, many halos have density contrasts well above 178 naively extrapolated from the top-hat model. At $z = 2.2$ there were 863 halos with mass larger than $1.5 \times 10^{11}$ $h^{-1}M_\odot$ and central mesh- defined overdensity larger than 50, which have actual overdensity larger than 178 for any of the radii defined above. Our maxima-on-mesh approach gave 775 halos, or 11% fewer. The difference is mainly due to the 20-25% difference in masses. We conclude that our estimates of number of halos are reasonably accurate. If anything, our simpler algorithm probably *underestimates* the number density of high-z objects.

For a filter $W$ with radius $r_f$ and mass $M = \alpha_m \rho_0 r_f^3$ ($\rho_0$ being the the critical density at present) the Press-Schechter approximation for the number density of halos with mass larger than $M$ at redshift $z$ is

$$N(>M,z) = \int_M^\infty n(m,z)dm = \sqrt{\frac{2}{\pi}} \frac{\delta_c}{\alpha_m} \int_{r_f}^\infty \frac{\epsilon(r_f,z)}{\sigma(r_f,z)} \exp\left(\frac{-\delta_c^2}{2\sigma(r_f,z)^2}\right) \frac{dr_f}{r_f^2} \;, \quad (1)$$

where

$$\epsilon(r_f, z) = \int k^4 P(k,z) \frac{W(kr_f)}{kr_f} \frac{dW(kr_f)}{d(kr_f)} dk \bigg/ \int k^2 P(k,z) W^2(kr_f) dk ,$$
$$\sigma(r_f, z) = \frac{1}{2\pi^2} \int k^4 P(k,z) W^2(kr_f) dk ,$$
(2)

and $P(k,z) = [(\Omega_c + \Omega_b)\sqrt{P_c} + \Omega_\nu \sqrt{P_\nu}]^2$ is the power spectrum of the total density. For a Gaussian filter $W = \exp[-(kr_f)^2/2]$ and $\alpha_m = (2\pi)^{3/2}$. For a top-hat filter $\alpha_m = 4\pi/3$.

The Press-Schechter approximation is commonly considered as a "good approximation." But in detail, it depends on how one uses it (e.g., Gaussian filter or top-hat filter) and what kinds of objects are considered. The approximation becomes very sensitive to the details once rare objects with a small fraction of the mass in them are considered. The reason is simple – we deal with the tail of a Gaussian distribution, where a small change in $\sigma$ or $\delta_c$ produces large variations in results. We regard $\delta_c$ as a parameter to be fit to the numbers of halos actually found in the simulations. Different people give different values for $\delta_c$. For example, for the CDM model and Gaussian filter Efstathiou & Rees (1988) gave $\delta_c = 1.33$; Carlberg & Couchman (1989) and Gelb (1992) gave $\delta_c = 1.44$. In principle, results can be different for different cosmological models.

Comparing to the KHPR93 and NKP94 simulations, we found that there is no unique value for the threshold parameter $\delta_c$ which can provide a good fit for all masses. The value of $\delta_c$ for a Gaussian filter for the Press-Schechter curve drawn for each mass is indicated in Figure 1(b). When finding dark halos in numerical simulations, we used a constant comoving sphere with effective radius of $0.18h^{-1}$Mpc. This is probably small for *large* objects with mass $\geq 10^{12}~M_\odot$. As the result, we miss a number of objects with this mass, which at least partially explains why the numerical results go below the Press-Schechter curves. But the objects that we missed are probably small groups of galaxies, not isolated galaxies. The large value of $\delta_c = 1.60$ found by KHPR93 was mainly based on simulations with very small box size $7h^{-1}$Mpc. Such a small box does not have large waves to produce large structures like superclusters or filaments, which should host high-redshift galaxies. Thus, it is quite likely that results from small boxes significantly underestimate the number of high redshift galaxies. Similarly, note that in Figure 1(b) the number densities from the smaller box are somewhat lower at higher redshift, although they agree at low redshift. This suggests that still larger boxes would give even higher number densities at high redshift. Predictions for Gaussian filter with $\delta_c = 1.33$ and $M = 1.5 \times 10^{11}~h^{-1} M_\odot$ are shown as the dotted curve. This curve goes above the results for the large box, but at $z > 2$ the difference is about the same as the difference between the results for small ($25h^{-1}$Mpc) and large ($50h^{-1}$Mpc) box simulations. It is quite reasonable to suggest that the numbers of halos with even larger boxes and better resolution would be close to $\delta_c = 1.33$ predictions.

This is also supported by our more accurate estimates of numbers of dark halos at $z = 2.2$. For the comparisons with data on damped Ly$\alpha$ systems in the next section, we therefore use $\delta_c = 1.33$, the same value used by Efstathiou & Rees (1988) for CDM. The top-hat model with traditional $\delta_c = 1.68$ (triangles) predicts systematically smaller number of dark dark halos at all redshifts. At $z < 1.5$ the difference, though visible, is not large. But it grows as we go to higher $z$. By $z = 3$ the top-hat model underestimates the number of halos by a factor of four compared to the results from the large-box simulation.

KHPR93 noted that the $\Omega_c/\Omega_\nu/\Omega_b = 0.6/0.3/0.1$ CHDM model predicted almost as many large galaxies at high redshifts as CDM with $b = 2.5$, probably enough to account for the numbers of bright quasars. The conclusions of Haehnelt (1993) and Liddle & Lyth (1993) are that this model may predict too few quasars if the number density of bright quasars at $z \approx 4$ is similar to that at $z = 2-3$. However, Warren, Hewett, & Osmer (1994) find a steep drop-off in the number density for redshifts beyond about 3.3, so the CHDM model may have no problem accounting for number densities of bright quasars.

### 3. Damped Ly$\alpha$ Systems

If $\Omega_b$ is the baryon density parameter and $f_g$ is the neutral fraction of the gas in the absorbers, then the density parameter contributed by matter collapsed in structures associated with such systems is

$$\Omega_{coll} = \frac{\Omega_g}{\Omega_b f_g}. \tag{3}$$

In the above relation, $\Omega_g$ is the total density parameter of neutral gas in damped Ly$\alpha$ systems, as provided by observational data (Lanzetta 1993, LWT93). In the following, we assume that $f_g = 1$, so that the neutral mass in a collapsed object of total mass $M$ is $\Omega_b M$. If CHDM cannot account for the damped Ly$\alpha$ observations with $f_g = 1$, it will have an even harder time if $f_g < 1$.

According to the Press-Schechter approximation, the fraction of the total volume, $F(M, z)$, associated with Gaussian–distributed fluctuations that collapse at redshift $z$ to structures of mass larger than $M$, is

$$F(M, z) = \text{erfc}\left(\frac{\delta_c}{\sqrt{2}\sigma(M, z)}\right), \tag{4}$$

where erfc($x$) is the complementary error function. Under the assumption of very small fluctuations at the time the collapse starts, eq. (4) coincides with the contribution to the density parameter at redshift $z$ due to structures of mass $> M$.

In Figure 2 we plot the resulting $\Omega_{coll}$ as a function of redshift for two different fluctuation spectra and at different mass scales. Figure 2(a) refers to the standard $\Omega_c/\Omega_\nu/\Omega_b =$ 0.6/0.3/0.1 CHDM model while Figure 2(b) is for the 0.675/0.25/0.075 model. This figure is to be compared with Figure 1 by Subramanian & Padmanabhan (1994). If we identify structures associated with Ly$\alpha$ absorbers as fluctuation with $\delta \geq \nu\sigma(r_f, z)$, then we will expect that, for fixed $\nu$, smaller scale fluctuations undergo non–linear evolution at higher redshifts. This is shown by the light dashed lines, which trace the redshifts at which fluctuations with $\nu = 1.5$ (upper line) and $\nu = 2$ (lower line) reach non–linearity (i.e., $\sigma(r_f, z) = \nu^{-1}$). For $M = 10^{12} M_\odot$ both models severely underestimate $\Omega_{coll}$ at high redshifts, while taking $M = 10^{10} M_\odot$ is consistent with observations up to $z \simeq 3$. The highest-redshift point is not reproduced by the 0.6/0.3/0.1 model, while the 0.675/0.25/0.075 model seems only marginally inconsistent with it.

We derive the relationship between the mass of a galaxy $M$, the comoving radius $r_f$ of the Gaussian filter and the circular velocity $V_c$ in the following simplified way. For the spherical collapse model (e.g. Peebles 1980), the proper radius $r_g$ of a galaxy collapsed at redshift $z$ is $r_g = r_f\sqrt{2\pi}(1+z)^{-1}[(4\pi/3)178]^{-1/3}$. Then $V_c \equiv (GM/r_g)^{1/2}$ and we get

$$M = \frac{V_c^3}{\sqrt{89}GH_0(1+z)^{3/2}} = 2.45 \times 10^{11} \ h^{-1} M_\odot \left(\frac{V_c}{100 \text{ km s}^{-1}}\right)^3 (1+z)^{-3/2}. \quad (5)$$

This is the mass we insert in the Press-Schechter expression (1).

In Figure 3 we compare the predictions of two CHDM models (full curves) for the total density of baryons in damped Ly$\alpha$ systems with the observational data (Lanzetta 1993, LWT93). We again assume the neutral fraction is unity. The predictions always rise with decreasing $z$ and provide only upper limits to the mass in neutral hydrogen in collapsed objects of various $V_c$. We find that if most of the observed Ly$\alpha$ absorption originates from galaxies with circular velocity larger than 100 km s$^{-1}$, the standard 0.6/0.3/0.1 CHDM model (left panel) has severe problems because it predicts too small a fraction of mass in these objects at redshifts $\geq 2.5$. However, if damped Ly$\alpha$ systems correspond to objects of circular velocity $\geq 50$ km s$^{-1}$, only the highest-redshift ($z = 3.0 - 3.4$) point is in disagreement with the predictions of the CHDM model. Because there is data on only four damped Ly$\alpha$ systems at $z > 3$ with no measurements of rotational velocities, it is difficult to say if the model is rejected or not. If, like quasars, the numbers of massive damped Ly$\alpha$ systems actually decline beyond $z = 3$ — for example, because the damped Ly$\alpha$ systems at higher redshift are mainly smaller-mass or still-collapsing systems (e.g., caustics) — then the model is still viable.

The dot-dash line in Figure 3(a) shows the result of an analogous Press-Schechter calculation using a top-hat filter and $\delta_c = 1.68$. This is in excellent agreement with the $V_c > 50$ km s$^{-1}$ curve in Figure 1 of Kauffmann & Charlot (1994), calculated under the same assumptions. This explains why we disagree with the conclusions of this and the other recent preprints that used a Press-Schechter approximation with these assumptions: they correspond to numbers of halos considerably smaller at high redshifts than indicated by the high-resolution N-body simulations of KHPR93 and NKP94.

The prediction for the 0.6/0.3/0.1 CHDM model is that most of the gas at redshifts beyond about 2.5 is in small clustered clumps. Note that those clumps are not necessarily each protogalaxies. It is quite likely that numerous small clumps (at $z = 3$ they have dark matter masses typically $\approx 5 \times 10^8\ h^{-1}M_\odot$, $V_c = 25$ km s$^{-1}$) can have a more massive neighbor — a forming galaxy, which could be observed at the redshift of the damped Ly$\alpha$ system. Later those clumps will merge and produce a real galaxy.

Predictions for numbers of objects at higher redshifts are exponentially sensitive to the fluctuation amplitude. Figure 3(b) shows that the $\Omega_c/\Omega_\nu/\Omega_b = 0.675/0.25/0.075$ variant of the CHDM model with a relatively small difference in parameters predicts significantly larger clumps. A typical object at $z = 3$ in this case has mass of $3.8 \times 10^9\ h^{-1}M_\odot$, radius $6.6 h^{-1}$kpc, and $V_c = 50$ km s$^{-1}$. This looks like a small (proto)galaxy. Now even the highest-$z$ point can be accounted for by halos with $V_c \sim 50$ km s$^{-1}$, and the question then arises what becomes of all this neutral hydrogen at lower redshifts.

We can make a more realistic (but more speculative) model for the damped Ly$\alpha$ systems if we suggest that it takes some time $\tau_{\text{gas}}$ for each galaxy to "digest" the gas – to convert most the gas into stars or to ionize it. In order to mimic this situation we estimate as before the density of baryons confined at redshift $z(t)$ in halos with rotational velocity larger than $V_c$, but assume that *all* the baryons in such halos at a time $\tau_{\text{gas}}$ earlier has now been ionized or converted to stars. Thus, we estimate the fraction of neutral gas at $z$ inside halos with $V > V_c$ is $\tilde{\Omega}_g(z) = \Omega_g(z) - \Omega_g(z(t - \tau_{\text{gas}}))$, where $\tau_{\text{gas}}$ is a characteristic life-time for the neutral gas. Like Kauffmann & Charlot (1994), we found that a constant $\tau_{\text{gas}}$ cannot reproduce the observed trend in $\Omega_g$. We need to assume that this process of "digesting" was less efficient in the past. Results for $\tilde{\Omega}_g(z)$ for a model with $\tau_{\text{gas}} = (1 + z)^2 0.35 \times 10^8$ yr are shown in Figure 3 as dashed curves.

## 4. Conclusions

The "standard" CHDM model with $\Omega_c/\Omega_\nu/\Omega_b = 0.6/0.3/0.1$ predicts a total density of neutral hydrogen in objects with circular velocity $V_c > 50$ km $s^{-1}$ that is consistent with the damped Ly$\alpha$ system observations reported in LWT93 at $z \leq 3$, but not with their highest-redshift data point. However, these predictions are quite sensitive to the parameters of the model, and for example in CHDM with 0.675/0.25/0.075, collapsed objects with $V_c \geq 50$ km $s^{-1}$ can account for all the data.

Thus it appears to be premature to conclude that CHDM is inconsistent with the damped Ly$\alpha$ data. However, it will be necessary to check by running new N-body simulations that CHDM models with parameters such as 0.675/0.25/0.075 will still have velocities on small scales that are compatible with the data. There is reason to expect that this will indeed be so, since the comparisons of 0.6/0.3/0.1 CHDM with the CfA data reported in NKP94 indicated that there may still be a little room available for lowering the neutrino mass (or, equivalently, $\Omega_\nu$). Such N-body simulations can also confirm that the numbers of halos are sufficiently accurately represented by the Press-Schechter approximation with a Gaussian filter and the parameters we have assumed. Another issue that needs to be checked is that for the range of neutrino mass considered here, the model does not overproduce clusters. (This is a concern, since it is well known that clusters are overproduced in the CDM limit $\Omega_\nu \longrightarrow 0$ if the fluctuation amplitude is normalized to COBE; cf. White, Efstathiou, & Frenk 1993. If producing the observed numbers of clusters requires that the fluctuation amplitude be lowered by 10%, for example, then we find that $\Omega_{coll}$ for $M = 10^{10} M_\odot$ decreases by a factor of about two at $z = 3$.) Eventually, with the increasing computational power afforded by massively parallel computers, it will be possible to extend such cosmological simulations by including hydrodynamics, energy input from stars and supernavae, and radiative transfer, to more realistically model the creation and evolution of structure including damped Ly$\alpha$ clouds.

However, it is already clear that in a model such as CHDM, in which most large structures form at relatively low redshifts, most damped Ly$\alpha$ systems at redshifts beyond about 2.5 must be associated with objects having relatively small circular velocities $V_c < 100$ km s$^{-1}$. Although it is impossible to estimate the circular velocities or velocity dispersions of such systems by observing the Ly$\alpha$ absorption itself (since its width is due to damping) such systems typically have substantial metallicity. The Doppler widths of the associated metal line systems or the relative velocities of different metal-rich clouds, which can be measured for example using the high-resolution spectrograph at the Keck telescope, should allow estimates of the velocities, and therefore the masses, of these objects. The clear prediction of CHDM is that these widths, relative velocities, and masses must decline with increasing redshift.

Acknowledgments. SB thanks UCSC for its hospitality during the first phase of preparation of this work. JRP acknowledges very helpful conversations with T. Walker and A. Wolfe, and support from NSF and the University of California, Santa Cruz. Simulations were done using the CONVEX-3880 at the NCSA.

FIGURE CAPTIONS

**Figure 1**. (a) Power spectra of fluctuations in the total density at the present epoch for CHDM and CDM with $\Omega = 1$. The lower three curves are for CHDM models with various fractions of baryons and hot dark matter (neutrinos). The lower full curve is for $\Omega_c/\Omega_\nu/\Omega_b = 0.60/0.30/0.10$. The dashed curve is for $0.655/0.27/0.075$. The dot-dashed curve is for $0.675/0.25/0.075$. The upper full curve is for the CDM model with the same normalization on very long waves.

(b) Number density of objects at different redshifts in the $0.60/0.30/0.10$ CHDM model. Results of numerical simulations are shown as dashed (KHPR93, box size $25h^{-1}$Mpc) and full (NKP94, box size $50h^{-1}$Mpc) curves. The dot-dashed curves indicate predictions based on the Press-Schechter approximation. The value of $\delta_c$ for a Gaussian filter for each mass is indicated in the Figure. The dotted curve is for a Gaussian filter with $\delta_c = 1.33$. Triangles show predictions for the top-hat filter with $\delta_c = 1.68$.

**Figure 2**. Fraction of the total density contributed by collapsed objects as a function of redshift. The data points are obtained from the observational data by Lanzetta et al. (1993) after suitably rescaling the total density in form of Ly$\alpha$ clouds (see text); indicated error bars correspond to $1\sigma$. Panel (a) is for the CHDM model with $\Omega_c/\Omega_\nu/\Omega_b = 0.60/0.30/0.10$, while panel (b) is for $0.675/0.25/0.075$. The light dotted lines trace the redshifts at which fluctuations of $1.5\sigma(r_f,z)$ and $2\sigma(r_f,z)$ (upper and lower lines, respectively) reach non–linearity. In all the cases, a Gaussian window is assumed with critical density contrast $\delta_c = 1.33$.

**Figure 3**. Fraction of the total density in the form of damped Ly$\alpha$ clouds: observational results of Lanzetta et al. (1993) compared with predictions of CHDM models with $\Omega_c/\Omega_\nu/\Omega_b$ given by (a) $0.60/0.30/0.10$, and (b) $0.675/0.25/0.075$. Error bars for the observational points correspond to $1\sigma$. The full curves indicate the fraction of baryons in collapsed objects with estimated circular velocities larger than $15, 25, 50, 100$ km s$^{-1}$ (from top to bottom), calculated using a Gaussian filter with $\delta_c = 1.33$. The curve for 15 km s$^{-1}$ is not shown on (b). The Press-Schechter predictions for a top-hat filter with $\delta_c = 1.68$ and circular velocity 50 km $s^{-1}$ are shown as the dot-dashed curve in (a). Results for a model of conversion of gas to stars with $\tau_{\text{gas}} = (1+z)^2 0.35 \times 10^8$ y are shown as dashed curves (see text).

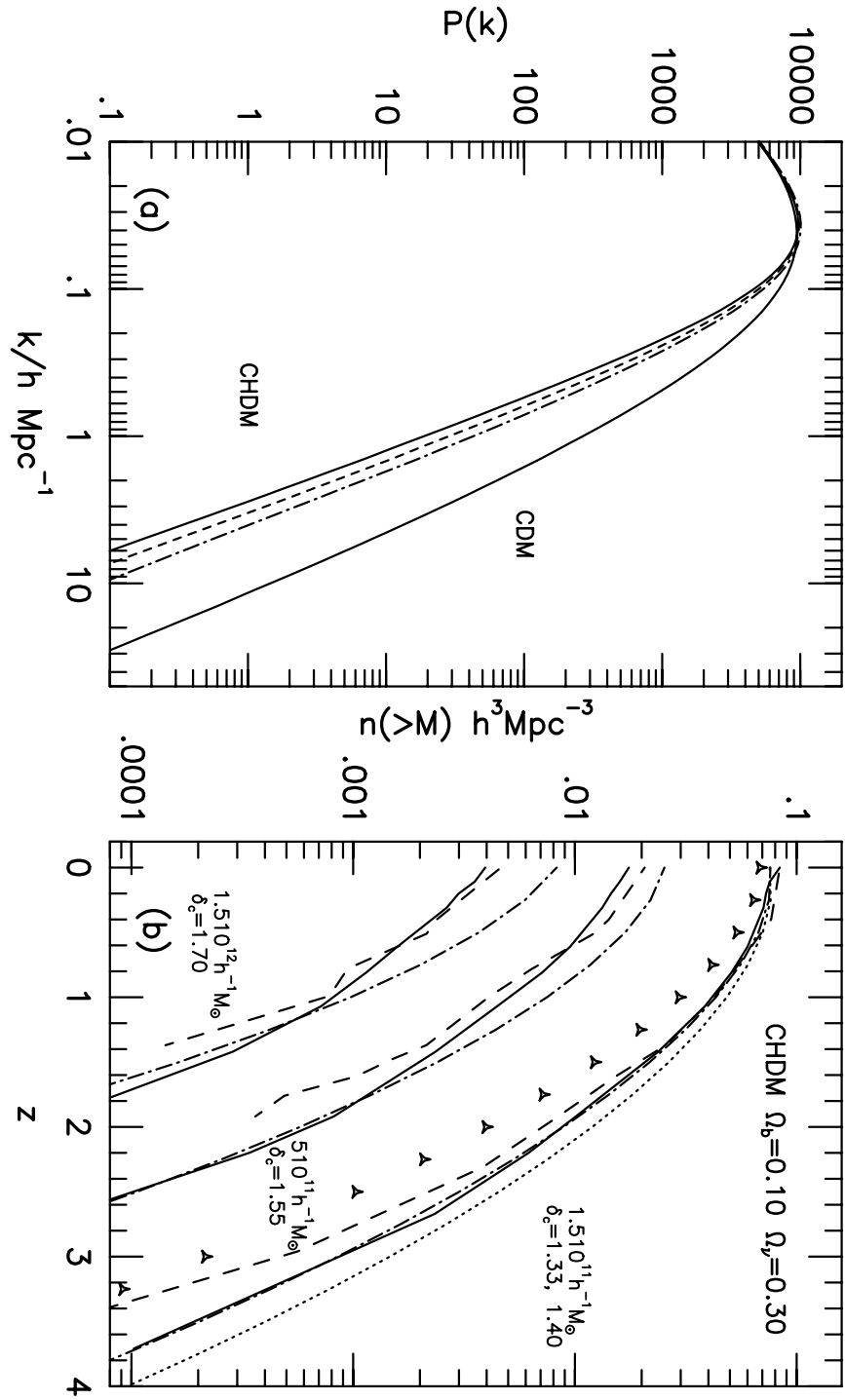

Figure 1

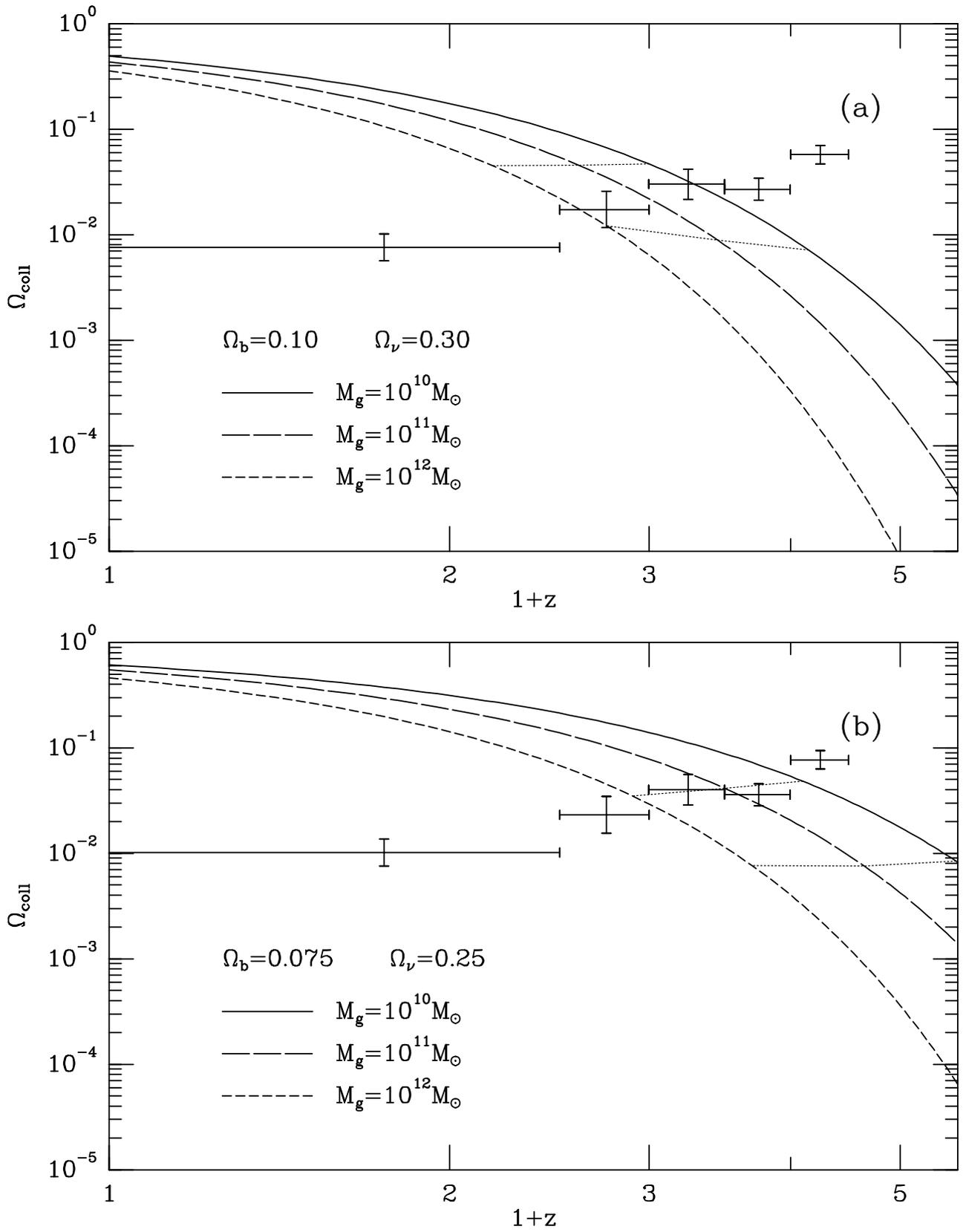

Figure 2

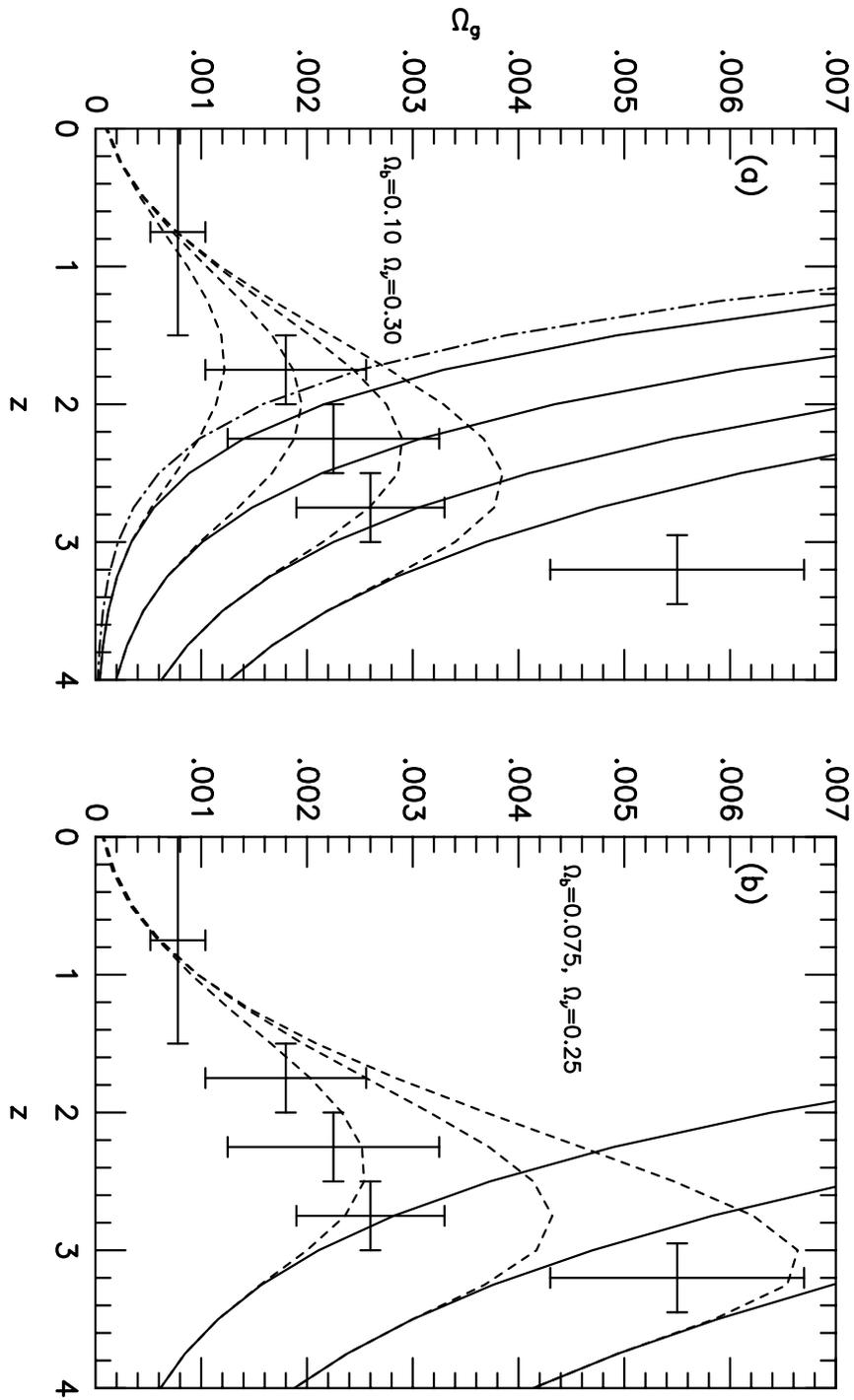

Figure 3